\documentclass[aps,prl,twocolumn,superscriptaddress,showpacs]{revtex4}
\usepackage{graphicx}
\begin{document}

\title{Design considerations for table-top, laser-based VUV and X-ray free electron lasers}

\author{F.~Gr\"{u}ner}
\email{florian.gruener@physik.uni-muenchen.de}
\affiliation{Department of Physics, Ludwig-Maximilians-Universit\"at M\"{u}nchen, 85748 Garching, Germany}
\author{S.~Becker}
\author{U.~Schramm}
\author{T.~Eichner}
\author{R.~Weingartner}
\author{D.~Habs}
\affiliation{Department of Physics, Ludwig-Maximilians-Universit\"at M\"{u}nchen, 85748 Garching, Germany}
\author{J.~Meyer-ter-Vehn}
\author{M.~Geissler}
\affiliation{Max-Planck Institute of Quantum Optics, 85748 Garching, Germany}
\author{M.~Ferrario}
\author{L.~Serafini}
\affiliation{INFN, 00044 Frascati, Italy}
\author{B.~van~der~Geer}
\affiliation{Pulsar Physics, 3762 Soest, The Netherlands}
\author{H.~Backe}
\author{W.~Lauth}
\affiliation{Institute of Nuclear Physics, Mainz, Germany}
\author{S.~Reiche}
\affiliation{University of California Los Angelos, UCLA, Los Angelos, USA}
\date{13.12.2006}

\begin{abstract}

A recent breakthrough in laser-plasma accelerators, based upon ultrashort high-intensity lasers,
demonstrated the generation of quasi-monoenergetic GeV-electrons. With future Petawatt lasers
ultra-high beam currents of $\sim100$ kA in $\sim10$ fs can be expected, allowing for drastic reduction
in the undulator length of free-electron-lasers (FELs). We present a discussion of the key aspects of a
{\em table-top} FEL design, including energy loss and chirps induced by space-charge and wakefields.
These effects become important for an optimized table-top FEL operation. A first proof-of-principle VUV
case is considered as well as a table-top X-ray-FEL which may open a brilliant light source also for
new ways in clinical diagnostics.

\end{abstract}

\maketitle

\section{Introduction} The pursuit of {\em table-top} FELs combines two rapidly developing fields:
laser-plasma accelerators, where the generation of intense quasi-monoenergetic electron bunches up to
the GeV range has been achieved \cite{nature1,nature2,nature3,gev}, and on the other hand large-scale
X-ray free-electron lasers (XFELs) that are expected to deliver photon beams with unprecedented peak
brilliance. A prominent application of such FEL pulses is single molecule imaging \cite{hajdu}. The
proposed laser-plasma accelerator-based FELs would not only allow a greater availability due to their
smaller size and costs, but also offer new features, such as pulses synchronized to the
phase-controlled few-cycle driver laser \cite{baltuska} for pump-probe experiments. Moreover, the X-ray
energy of a table-top XFEL can be as large as required for medical diagnostics (above $20$ keV
\cite{bravin}).

The mechanism for the generation of intense (nC charge) quasi-monoenergetic electron pulses by laser-plasma
accelerators requires an ultrashort, high-intensity laser-pulse with a length shorter than the plasma
wavelength (on the $\mu$m-scale corresponding to gas densities of $10^{19}$ cm$^{-3}$). Due to the
ponderomotive force, plasma electrons are blown out transversely, leaving an electron-free zone - the {\em
bubble} - behind the laser pulse \cite{bubble}. These electrons return to the axis after half a plasma
oscillation, thus determining the size of the bubble in the order of the plasma wavelength. Typically about
$10^{9}...10^{10}$ electrons are captured into the bubble, as found both experimentally and from scaling laws
of relevant bubble parameters within the framework of similarity arguments \cite{pukhov_scaling}. Due to the
inertial positive ion background, these electrons experience a strong electrical field gradient of up to TV/m.
Particle-in-cell (PIC) simulations \cite{geissler} show that the bubble electrons form a stem that is
geometrically considerably smaller than the bubble as seen in Fig. 1. 

\begin{figure}[htb]
\centering
\includegraphics*[width=80mm,angle=-90]{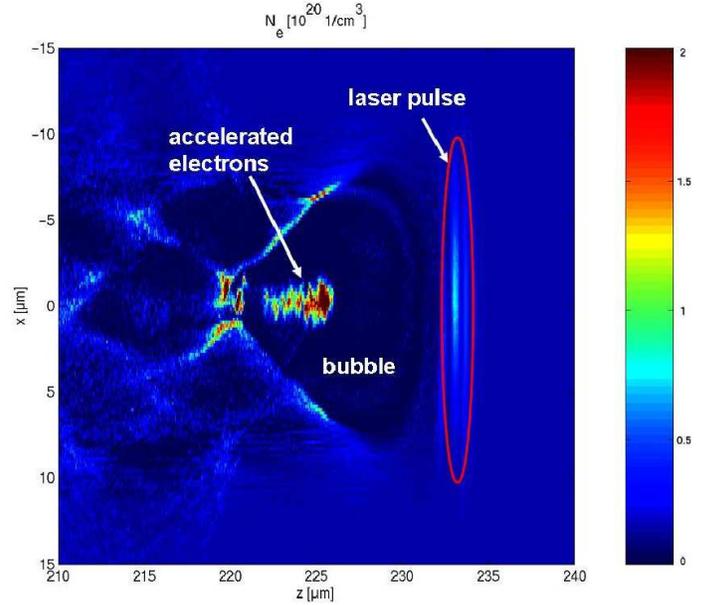}
\caption{(Color online) Snapshot of electron density (in units of $10^{20}$ cm$^{-3}$) from PIC 
simulation. The geometrical
size of the electron-free cavity (``bubble") behind the laser spot corresponds to the plasma period, 
which is about $8$ $\mu$m in this case 
(gas density $1.8\cdot 10^{19}$ cm$^{-3}$). 
The stem of the
high-energy electrons is much shorter than the plasma period.}
\label{fig:1}       
\end{figure}

For these dimensions, beam currents of the order of $\sim100$ kA (i.e. $\sim1$ nC charge within $\sim10$ fs)
can be reached without the need for any bunch compressor.\\ For preparation of the required gas densities,
either supersonic gas-jets or capillaries \cite{hooker} are used, the latter consisting of a
$\sim300~\mu$m-thin gas-filled channel with a parabolic radial ion density profile generated by a second laser
pulse or discharge. The advantage of the capillary is that the laser can be guided beyond the Rayleigh length
allowing longer acceleration distances and hence higher electron energies. The very recent experiments by
Leemans et al. \cite{gev}, utilizing a $40-$TW laser pulse of $38$ fs duration and a gas density of $4.3\cdot
10^{18}$ cm$^{-3}$, clearly show that $1$ GeV electron beams can now be produced with capillaries. However,
the measured charge of $30$ pC is significantly below our design goal of $1$ nC. In order to further improve
the resulting current, we propose the scheme described below. The number of bubble electrons can be increased
when using plasmas with higher density such as in Fig. 1, because the feeding process is more efficient if
more plasma electrons are present. An increased gas density requires shorter laser-pulses (sub-$10$-fs), such
that the entire laser pulse fits into one plasma period. In this case, the entire laser pulse energy can be
used, while longer pulses lose energy during self-shortening for entering the bubble regime. Therefore, in
contrast to all previous laser-acceleration experiments, we propose to use a different driver laser, namely a
$\sim5$ fs-Petawatt laser (like the Petawatt-Field-Synthesizer PFS at MPQ \cite{pfs}). Own PIC simulations
show that such ultrashort lasers can capture more than $1$ nC charge inside the bubble. The smaller plasma
period leads to smaller bubbles, hence shorter bubble stems and thus higher currents ($1$ nC in $10$ fs).
However, the expected final energy in the bubble regime is, according to the scaling laws, lower for shorter
laser pulses. To overcome this limit, we also propose to use a capillary and suggest the following scenario
therein: with a longitudinal plasma density gradient, the laser pulse can be forced to gradually increase its
diameter, thus adiabatically turning from the pure bubble regime into the pure blow-out regime long before the
laser is depleted. Fig. 1 shows a snapshot of a bubble around the transition from the bubble to the blow-out
regime, where the electrons blown away from the laser and flowing around the bubble are so strongly deflected
by the electric field of the captured bubble electons, that electrons are no more scattered into the bubble
(which is not charge-neutralized yet). The number and absolute energy spread of the bubble electrons is thus
frozen, but their energy is still increased due to the present bubble fields. The remaining energy of the
laser allows to maintain the bubble structure for the remaining distance inside the capillary, where the laser
is then guided along to overcome the Rayleigh limit. Only for this stage with increased laser beam size, the
capillary-induced guiding is relevant. Note that this scheme is a two-stage approach, but within one and the
same capillary, where the laser turns adiabatically from one stage into the other. It can be expected that the
energy spread for $100$ MeV is about $1~\%$ (as confirmed by measurements \cite{nature2}), but $0.1\%$ for $1$
GeV \cite{note}. Normalized emittances are found both from simulations and experiments to range between
$0.1-1$ mm$\cdot$mrad.

\section{Table-top Free-Electron-Lasers}
An FEL requires an undulator which is an arrangement of magnets with an alternating transverse magnetic
field. Electrons in an undulator are forced on a sinusoidal trajectory and can thus couple with a
co-propagating radiation field. The induced energy modulation yields a current modulation from the
dispersion of the undulator field. This modulation is called micro-bunching expressing the fact that
the electrons are grouped into small bunches separated by a fixed distance. Therefore, electrons emit
{\em coherent} radiation with a wavelength equal to the periodic length between the micro-bunches. In a
{\em Self-Amplification of Spontaneous Emission} (SASE) FEL, there is no initial radiation field and the
seed has to be built up by the spontaneous (incoherent) emission \cite{sase}.\\ We present quantitative
arguments, which are complemented by SASE FEL ({\em GENESIS 1.3} \cite{genesis}) simulations.  The gain
length, which is the e-folding length of the exponential amplification of the radiation power, is

\begin{equation}
\label{lgain}
L_{gain,ideal}=\frac{\lambda_{u}}{4\pi\sqrt{3}\rho},
\end{equation}

with undulator period $\lambda_{u}$ and the basic scaling parameter $\rho$. This Pierce or FEL parameter
describes the conversion efficiency from the electron beam power into the FEL radiation power and reads for the
one-dimensional and {\em ideal} case (neglecting energy spread, emittance, diffraction, and time-dependence)
\cite{sase,xie}

\begin{equation}
\label{rho}
\rho = \frac{1}{2\gamma}\left[ \frac{I}{I_{A}}\left( \frac{\lambda_{u}A_{u}}{2\pi
\sigma_{x}}\right)^{2}\right]^{1/3}.
\end{equation}

Here $\gamma = E_{beam} / mc^{2}$ is the electron beam energy, $I$ the beam current, $I_{A}=17 kA$ the
Alfven-current, $\sigma_{x}$ the beam size and $A_{u}=a_{u}[J_{0}(\zeta)-J_{1}(\zeta)]$ (planar undulator),
whereby $a_{u}^{2}=K^{2}/2$, $K$ the undulator parameter ($K=0.93\cdot \lambda_{u}[cm]\cdot B_{0}[T]$ and
$B_{0}$ the magnetic field strength on the  undulator axis), $\zeta=a_{u}^{2}/(2(1+a_{u}^{2}))$, and $J$ are
Bessel functions. In presence of energy spread, emittance, and diffraction, a correction factor $\Lambda$ is
introduced for the gain length \cite{xie}: $ L_{gain}=L_{gain,ideal} (1+\Lambda)$. The saturation length
$L_{sat}$ is the length along the undulator at which maximum micro-bunching is reached. The power of the FEL
radiation at this point is the saturation power $P_{sat}$, while the shot noise power $P_{n}$ is the power of
the spontaneous undulator radiation from which the SASE FEL starts. The coupling factor $\alpha=1/9$ describes
the efficiency at which the noise power couples to the FEL gain process. The saturation length reads

\begin{equation}
\label{lsat}
L_{sat}=L_{gain}\cdot ln(\frac{P_{sat}}{\alpha \cdot P_n}),
\end{equation}

where the saturation power scales as

\begin{equation}
\label{psat}
P_{sat}\sim \left( \frac{1}{1+\Lambda} \right)^{2}(I\cdot \lambda_{u})^{4/3}.
\end{equation}

The FEL-wavelength is, in case of a planar undulator,

\begin{equation}
\label{lambda}
\lambda=\frac{\lambda_{u}}{2\gamma^{2}} \left[ 1+\frac{K^{2}}{2} \right].
\end{equation}

Thus, for reaching a certain wavelength $\lambda$, a shorter undulator period $\lambda_{u}$ allows using less
energetic electrons. In the following Tab. 1 and Fig. 2, we compare the  {\em FLASH} VUV FEL (DESY)  in the
so-called {\em femtosecond mode} \cite{flash} with our corresponding {\em table-top} VUV FEL as well as a
table-top XFEL operating with $1.2$ GeV electrons, discussed further below.

\begin{table}[hbt]
\begin{center}
\caption{Parameters for the comparison between the DESY femtosecond-mode {\em Flash}-VUV case and the corresponding
table-top VUV FEL as well as a table-top X-ray FEL (rms values).}
\begin{tabular}{|l|c|c|c|}
\hline
\textbf{Parameter} & \textbf{FLASH (fs)} & \textbf{TT-VUV-FEL} & \textbf{TT-XFEL} \\ \hline
current & 1.3 kA & 50 kA  & 160 kA \\
norm. emitt. & 6 mm$\cdot$ mrad &  1 mm$\cdot$ mrad  & 1 mm$\cdot$ mrad\\
beam size & 170 $\mu$m & 30 $\mu$m & 30 $\mu$m\\
energy & 461.5 MeV & 150 MeV   & 1.74 GeV \\
energy spread  & 0.04 \% & 0.5 \%  & 0.1 \%\\
und. period & 27.3 mm & 5 mm   & 5 mm \\
wavelength & 30 nm & 32 nm   & 0.25 nm \\
Pierce par. & 0.002 & 0.01   & 0.0015 \\
sat. length & 19 m & 0.8 m   & 5 m\\ 
pulse length & 55 fs & 4 fs   & 4 fs\\ 
sat. power & 0.8 GW & 2.0 GW   & 58 GW\\ \hline
\end{tabular}
\label{l2ea4-t1}
\end{center}
\end{table}

\begin{figure}[htb]
\centering
\includegraphics*[width=80mm]{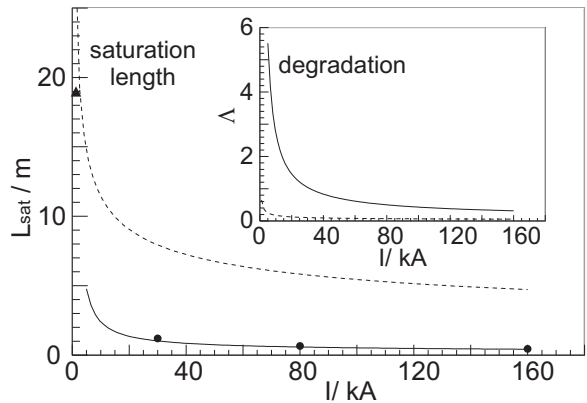}
\caption{Saturation lengths (Eq. \ref{lsat}) and (inset) degradation factor $\Lambda$ as a function of electron beam current $I$: DESY's {\em FLASH} (dashed curve, triangle) and the
table-top VUV scenario (solid line). The circles denote specific {\em GENESIS} runs.}
\label{fig:2}       
\end{figure}

The importance of the ultra-high current in the table-top case is evident. The smaller undulator period
allows a smaller beam energy, hence decreasing the gain and saturation lengths. The Pierce parameter gives
the upper limit of the acceptable energy spread $\sigma_{\gamma}/\gamma$. Thus, for compensating the
relatively large energy spread of laser-plasma accelerators and for maintaining a large output power, a beam
current significantly above $\sim17$ kA is mandatory for keeping $\rho$ large and $\Lambda$ small. 

\subsection{Space charge effects} Such ultra-high beam currents are subject to strong space-charge
forces. After release into vacuum, the electron bunch starts expanding, for which there are two sources:
(i) its own space-charge, driving a Coulomb-explosion, i.e. (transverse) space-charge expansion and
(longitudinal) debunching, and (ii) its initial divergence, which drives a linear transverse expansion. Such
expansions transform potential energy into kinetic, hence changing the initial energy distribution and
bunch form. First, we want to discuss an extreme case, that is, an upper limit of the charge that can
be expected from a laser-plasma accelerator at lower energies. For this study we take a beam energy of
$\gamma_{0}=260$, $1.25$ nC charge, and a Gaussian bunch with sizes  $\sigma_{x,y,z}=1\mu$m,
corresponding to $I=150$ kA, and an initial divergence of $\theta_{0}=1$ mrad. In the rest frame of the
bunch, its length amounts to $\sigma_{z}'=\sigma_{z}\cdot\gamma_{0}=260~\mu$m. For such an aspect-ratio,
the transverse  Coulomb-explosion dominates over the longitudinal one (GPT \cite{gpt}  simulations reveal
that after a time period of $1$ ps $\sigma_{x}'=73~\mu$m, while $\sigma_{z}'=268~\mu$m).  Fig. 3 shows
the electron distribution in the bunch rest frame from the GPT simulations.  

\begin{figure}[htb]
\centering
\includegraphics*[width=30mm,angle=-90]{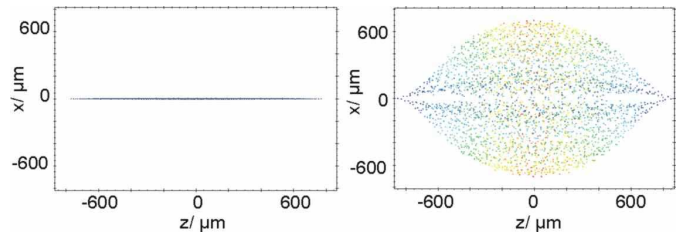} 
\caption{(Color online) Projected spatial electron distribution within the bunch rest frame in the extreme
case, at $\tau=0$ s (left) and $\tau=4$ ps (right), from a GPT run. The color code indicates the electron
energy (blue is initial kinetic energy, red is increased energy). The transverse expansion is clearly
dominating.} 
\label{fig:3}
\end{figure}

It can be seen that the transverse Coulomb explosion dominates the longitudinal one. We want to emphasize here
that studies of space-charge effects of such ultra-dense relativistic bunches can be simulated most accurately
only with codes (such as {\em GPT}  and {\em CSRtrack} \cite{csrtrack}) which utilize point-to-point
interactions (a result also found in Ref. \cite{leemans_spacecharge}). In contrast, other codes merely based
upon Poisson-solvers cannot cope with large relative particle motions within the bunch rest frame in case of
large initial energy spreads and Coulomb explosion-induced motion.\\ In comparison with the Coulomb-driven
explosion, the linear divergence-driven transverse expansion is even larger: after a distance of $1$ cm in the
laboratory frame, the beam size is increased to $10~\mu$m, while transverse Coulomb-explosion yields $6~\mu$m
only. Hence, even in the extreme case the transverse bunch expansion is dominated by the initial divergence
and for lower charges this dominance increases. If $\gamma_{0}$ is the bunch energy just before release into
vacuum, $\gamma'$ and $\beta'$ the energy and velocity of a test electron in the bunch rest frame, $\gamma$
and $\beta\approx 1$ its laboratory frame energy and velocity, then one can clearly distinguish between a
transverse-dominated expansion and a longitudinal-dominated one: in the first case, where the electron moves
in the rest frame with velocity $\pm\beta^{'}$ along transverse direction, $\gamma=\gamma_{0}\cdot\gamma'$,
while in the latter case, where the electron moves with velocity $\pm\beta'$ in beam direction,
$\gamma=\gamma_{0}\cdot\gamma'\cdot(1\pm\beta')$. Hence, in a purely transverse expansion {\em all} electrons
gain kinetic energy, while in a purely longitudinal expansion, some electrons lose energy. The inset of Fig. 4
shows the absolute kinetic electron energy distribution along the bunch at a distance of $z=3$ cm after its
release into vacuum (in case of $0.6$ nC). This spectrum shows that the bunch under study undergoes mostly a
transverse expansion. The maximum energy gain lies in the middle of the bunch, where the density is highest,
i.e., where initially the highest potential energy was stored. That the space-charge driven transverse
expansion is weaker than the divergence-driven one is also confirmed by the fact that (even in the extreme
case) the divergence is increased only to $1.3$ mrad. After passing a focusing system (a triplet of
quadrupoles, positioned at $z=4$ cm, having a length of about $10$ cm), the Gaussian-like energy distribution
is transformed into a quasi-linear one, as depicted in Fig. 4. The reason is that in the resulting
quasi-collimated low-divergence beam faster electrons can catch up the slower ones. Note that after $z=3$ cm
(inset) the fastest electrons are symmetrically distributed around the {\em middle} of the bunch. These
electrons start overtaking the slower ones, that is, they move forward within the bunch. If initially
longitudinal debunching dominates, the fastest electrons are found in the head and the slowest ones in the
tail of the bunch. 

\begin{figure}[htb]
\centering
\includegraphics*[width=80mm]{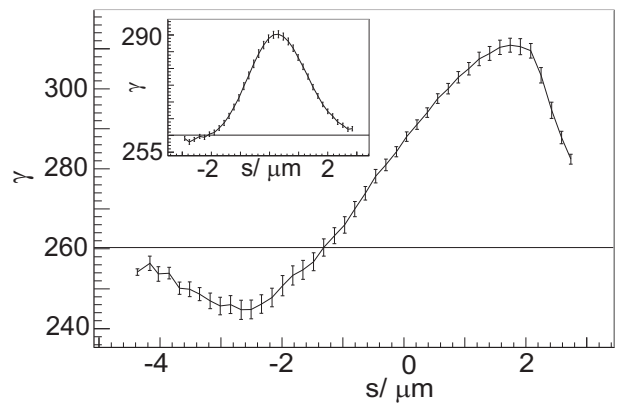}
\caption{Energy vs. position (in co-moving frame) along the bunch before entering the undulator (GPT
simulation): the Gaussian-like distribution
after a drift of $z=3$ cm behind the gas-jet (inset), indicating a transverse-dominated bunch
expansion, is stretched into a quasi-linear one behind the focusing system. The vertical lines show the slice energy spread 
$\sigma_{\gamma}$ and the horizontal lines mark the initial energy $\gamma_{0}=260$.}
\label{fig:4}       
\end{figure}

A key parameter for SASE to occur is the energy spread $\sigma_{\gamma}/\gamma$, which must always be smaller
than the Pierce parameter $\rho$. The relevant energy spread is taken over a bunch slice with the size of one
cooperation length, i.e. the slippage length of the radiation along the bunch over one gain length. As seen in
Fig. 4, a large fraction of the bunch fulfills $\sigma_{\gamma}/\gamma<\rho$. Due to the slippage of the
radiation relative to the bunch, the linear energy chirp must be compensated, because the energy factor $\gamma$
in Eq. (\ref{lambda}) increases along the bunch. However, in case of short-period undulators the gap is
correspondingly small (typically one third of the undulator period) and such small gaps lead to strong
wakefields also causing an energy variation along the bunch \cite{wakefield}. In case of ultrashort bunches, the
characteristic length of the resistive wake potential is longer than the bunch length and, hence, the energy
change is found to be negative with a linear few-percent variation along the bunch. This in turn implies that
the bunch as a whole is decelerated during the passage of the undulator, whereby the head loses less energy
than the tail. It turns out that the linear energy chirp induced by space-charge corresponds well with the
wakefield-induced slowing down of the entire electron bunch. Therefore, the radiation slipping in forward
direction interacts with electrons of effectively constant energy, if by varying the gap the bunch energy loss
is tuned with respect to the slippage. In other words, the space-charge effects and the wakefield effects
cancel each other. We have found with simulations that for a first proof-of-principle  of a table-top  FEL the
comparably low energy of $130-150$ MeV is well suited to cope with the wakefields in the undulator. In this
sense the table-top VUV case given in Tab. 1 can be regarded as an optimal test case.\\   Coherent synchrotron
radiation (CSR) \cite{csr} as another source for increasing the (slice) energy spread within the undulator was
found to have negligible impact. This was shown with {\em CSRtrack} simulations which take into account that in
our cases the bunches have a much larger transverse size ($\sigma_{x,y}\sim30~\mu$m) than length
($\sigma_{z}\sim1~\mu$m).\\ 

\subsection{Table-top XFEL} So far we have discussed a proof-of-principle scenario at relatively low electron
energies, where space-charge effects play a dominant role leading to a linear energy chirp. The situation of
the proposed table-top X-FEL (TT-XFEL) is different. For reaching a wavelength of $\lambda=0.25$ nm, an
electron energy of $1.74$ GeV is needed in case of a period of $\lambda_{u}=5$ mm. Space charge effects are
much weaker here. For this less demanding situation we have confirmed with four different simulation codes
({\em GPT, ASTRA \cite{astra}, CSRtrack, HOMDYN \cite{homdyn}}) that above $1$ GeV, with the same parameters
as in the extreme case given above, Coulomb-explosion leads to a projected energy chirp of below $0.3\%$ and a
bunch elongation with a factor below $1.1$. Furthermore, CSR can be neglected in agreement with the 1D theory.
Experimentally the most demanding constraint is that the Pierce parameter is one order of magnitude smaller
than in the test case at $130$ MeV, hence the electron (slice) energy spread should be as small as $0.1\%$.
This goal seems to be within reach considering the bubble-transformation scenario mentioned above. \\ Without
the effect of wakefields {\em GENESIS} simulations have shown that this TT-XFEL scenario with an undulator
length of only $5$ m yields $8\cdot 10^{11}$ photons/bunch  within $\sim4$ fs and $0.2\%$ bandwidth, a
divergence of $10~\mu$rad, and a beam size of $20~\mu$m. However, the wakefields become the dominant degrading
effect as the required undulator length is larger and the Pierce parameter reduced, hence also the tolerance
with respect to variations in the electron energy. But since there is no initial space-charge-induced energy
chirp, one must find another method for compensating the wakefield-induced energy variation. A suitable method
for compensating the wakefields for the TT-XFEL would be tapering, i.e., varying the undulator period along
the undulator. Due to the fact that the undulator parameter $K$ is smaller than unity, tapering via $K$, i.e.,
by gap variation, could only be used as fine-tuning. Depending on the specific material properties of the
undulator surface, our first wake calculations show that the bunch center loses over the entire $5$ m
undulator length about $10\%$ of its initial energy. Due to the linear wake field variation along the bunch
only a fraction of it will undergo SASE. A detailed study will be further investigated in a future paper.

\subsection{Experimental Status}
In order to demonstrate {\em practical} feasibility, we have built and tested a miniature focusing
system and undulator consisting of permanent magnets. The focusing triplet consists of mini-quadrupoles
with an aperture of just $5$ mm, hence  allowing for measured gradients of $530$ T/m. Their focusing
strength was measured with a $600$ MeV electron beam (at Mainz Microtron facility MAMI, Mainz,
Germany). A first test hybrid undulator with a period of only $5$ mm and a peak magnetic field strength
of $\sim1$ T has been built and produced with a $855$ MeV beam (also at MAMI) an undulator radiation spectrum
as expected. These results will be published elsewhere. 

\section{Conclusion}
We have shown by means of analytical estimates and SASE FEL simulations that
laser-plasma accelerator-based FELs can be operated with only meter-scale undulators. The key
parameter is the ultra-high electron peak current, which significantly reduces the gain length and
increases the tolerance with respect to energy spread and emittance. The latter would also allow to
increase the TT-XFEL photon energy into the medically relevant range of $20-50$ keV, because the
limiting quantum fluctuations in the spontaneous undulator radiation scale with $\gamma^{4}$
\cite{quantum_fluctuation} and the required electron energy of the TT-XFEL is comparatively small.

We thank W. Decking, M. Dohlus, K. Fl\"ottmann, T. Limberg, and J. Rossbach (all DESY) for fruitful
discussions. Supported by Deutsche Forschungsgemeinschaft through the DFG-Cluster of Excellence
Munich-Centre for Advanced Photonics. This work was also supported by DFG under Contract No. TR18.


\begin{thebibliography}{10}
\bibitem{nature1}
S.P.D. Mangles {\em et al.}, Nature {\bf 431}, 535 (2004)
\bibitem{nature2}
C.G.R. Geddes {\em et al.}, Nature {\bf 431}, 538 (2004) 
\bibitem{nature3}
J. Faure {\em et al.},Nature {\bf 431}, 541 (2004) 
\bibitem{gev}
W.P. Leemans {\em et al.}, Nature Physics {\bf 2}, 696 (2006) 
\bibitem{hajdu}
R. Neutze {\em et al.}, Nature {\bf 406}, 752 (2000) 
\bibitem{baltuska}
A. Baltuska {\em et al.}, Nature {\bf 421}, 611 (2003)
\bibitem{bravin}
W. Thomlinson, P. Suortti, and D. Chapman, Nucl. Instrum. Methods Phys. Res. A {\bf 543}, 288
(2005)
\bibitem{bubble}
A. Pukhov and J. Meyer-ter-Vehn, Appl. Phys. B {\bf 74}, 355 (2002)
\bibitem{pukhov_scaling}
A. Pukhov and S. Gordienko, Phil. Trans. R. Soc. A {\bf 364}, 623 (2006)
\bibitem{geissler}
M. Geissler, J. Schreiber, and J. Meyer-ter-Vehn, New J. Phys. {\bf 8} 186 (2006)
\bibitem{hooker}
D.J. Spence, A. Butler, and S. Hooker, J. Opt. Soc. Am. B {\bf 20}, 138 (2003) 
\bibitem{pfs}
http://www.attoworld.de/research/PFS.html
\bibitem{note}
Note that the detector resolution in the cited GeV-experiment did not suffice to resolve
energy spreads below one percent
\bibitem{sase}
R. Bonifacio, C. Pellegrini, and L.M. Narduci, Opt. Comm. {\bf 50}, 373 (1984)
\bibitem{genesis}
S. Reiche, Nucl. Instrum. Methods Phys. Res. A {\bf 429}, 243 (1999)
\bibitem{xie}
M. Xie, Nucl. Instrum. Methods Phys. Res. A {\bf 445}, 59 (2000)
\bibitem{flash}
Technical Design Report: http://www-hasylab.desy.de/facility/fel/vuv/
\bibitem{gpt}
M.J. de Loos, S.B. van der Geer, Proc. 5th Eur. Part. Acc. Conf., Sitges, 1241 (1996)
\bibitem{csrtrack}
M. Dohlus and T. Limberg, Proceedings of the FEL 2004 Conference, 18-21
\bibitem{leemans_spacecharge}
G. Fubiana, J. Qiang, E. Esarey, and W.P. Leemans, Phys. Rev. ST Accel. Beams {\bf 9}, 064402 (2006)
\bibitem{csr}
E. Saldin, E.A. Schneidmiller, and M.V. Yurkov, Nucl. Instrum. Methods Phys. Res. A {\bf 417}, 158
(1998)
\bibitem{astra}
K. Floettmann, Astra User's Manual, 2000, http://www.desy.de/~mpyflo/
\bibitem{homdyn}
M. Ferrario, J. E. Clendenin, D. T. Palmer, J. B. Rosenzweig, L. Serafini, Proc. of the 2nd ICFA Adv. Acc.
Workshop on ``The Physics of High Brightness Beams", UCLA, Nov., 1999; see also SLAC-PUB-8400 
\bibitem{wakefield}
A.W. Chao, {\em Physics of collective beam instabilities in high energy accelerators} (John Wiley \&
Sons, New York, 1993); K. Bane, SLAC-PUB-11829 (2006)
\bibitem{quantum_fluctuation}
M. Sands, Phys. Rev. {\bf 97}, 470 (1955)

\end{thebibliography}
\end{document}